\documentclass[11pt,a4paper]{article}
\usepackage{amssymb}
\usepackage{amsthm}
\usepackage{amsmath}
\usepackage{graphicx}
\usepackage{xcolor}
\usepackage{subcaption}
\usepackage{enumitem}
\usepackage{lineno}
\usepackage{comment}
\usepackage{algorithm}
\usepackage[indLines=false] {algpseudocodex}
\usepackage{hyperref}
\usepackage{authblk}
\hypersetup{
    colorlinks = true,
    linkcolor = blue,
    anchorcolor =blue,
    citecolor = blue,
    filecolor = blue,
    urlcolor = blue,
    pdfauthor=author
}

\def\O{\mathcal{O}}

\def\CH{\mbox{{\it CH}}}
\def\RH{\mbox{{\it RH}}}

\def\NW{\mbox{{\it NW}}}
\def\NE{\mbox{{\it NE}}}
\def\SW{\mbox{{\it SW}}}
\def\SE{\mbox{{\it SE}}}

\def\min{\mbox{{\it min}}}
\def\max{\mbox{{\it max}}}

\newcounter{problem}
\newtheorem{theorem}{Theorem}
\newtheorem{lemma}[theorem]{Lemma}

\newtheorem{prob}[problem]{Problem}

\title{An e{f}{f}icient algorithm for identifying rainbow ortho-convex 4-sets in \texorpdfstring{$k$}{k}-colored point sets}
\date{}

\author[1]{David Flores-Pe\~naloza\thanks{Email: dflorespenaloza@ciencias.unam.mx}}

\author[2]{Mario A. Lopez\thanks{Email: Mario.Lopez@du.edu}}

\author[1]{Nestaly~Mar\'in\thanks{Email: nestaly@ciencias.unam.mx}}

\author[3]{David Orden\thanks{Email: david.orden@uah.es}}

\affil[1]{Departamento de Matem\'aticas, Facultad de Ciencias, Universidad Nacional Aut\'onoma de M\'exico, Mexico}

\affil[2]{Department of Computer Science, University of Denver, US}

\affil[3]{Departamento de F{\'{\i}}sica y Matem\'aticas, Universidad de Alcal\'a, Spain}

\begin{document}
\maketitle

\begin{abstract}
Let $P$ be a $k$-colored set of $n$ points in the plane, $4 \leq k \leq n$.
We study the problem of deciding if $P$ contains a subset of four points of di{f}{f}erent colors such that its Rectilinear Convex Hull has positive area. We show this problem to be equivalent to deciding if there exists a point $c$ in the plane such that each of the open quadrants de{f}ined by~$c$ contains a point of $P$, each of them having a di{f}{f}erent color.
We provide an $O(n \log n)$-time algorithm for this problem, where the hidden constant does not depend on $k$; then, we prove that this problem has time complexity $\Omega(n \log n)$ in the algebraic computation tree model. No general position assumptions for $P$ are required.
\end{abstract}



\section{Introduction}

In the classical setting, a set in the Euclidean plane is said to be \emph{convex} if its intersection with any line is either empty or connected.
The \emph{convex hull} of a geometric set $S$, denoted as $\CH(S)$, is the intersection of all the closed half-planes containing~$S$.
Equivalently, $\CH(S)$ is the closed region obtained by removing from~$\mathbb{R}^2$ all the open half-planes that do not contain points of~$S$.

\begin{figure}[ht!]
	\centering
	\begin{subfigure}[t]{0.3\linewidth}
		\includegraphics[width=\linewidth] {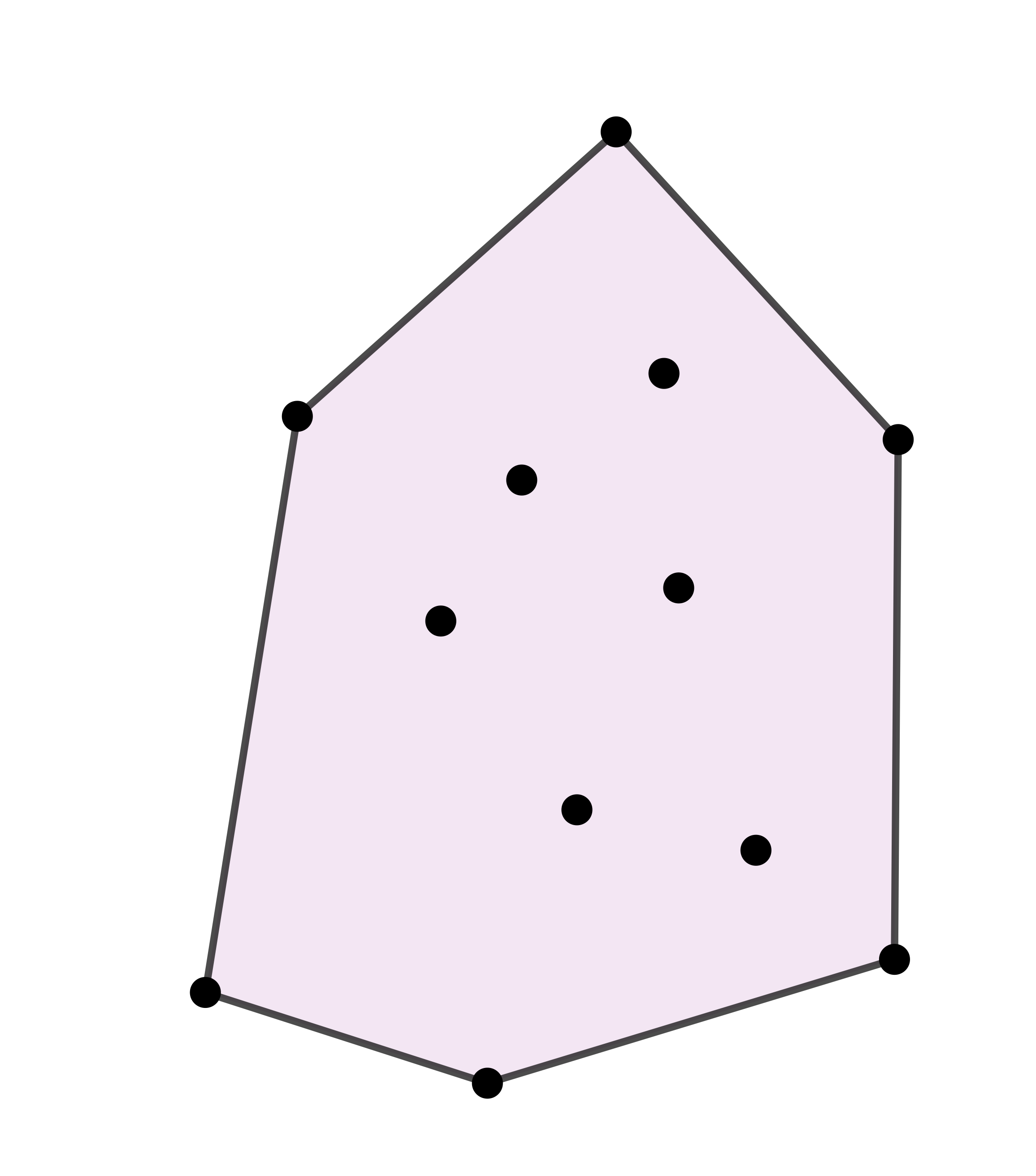}
		\caption{}
		\label{fig_ccc_ohull_1}
	\end{subfigure}
	~
	\begin{subfigure}[t]{0.3\linewidth}
		\includegraphics[width=\linewidth] {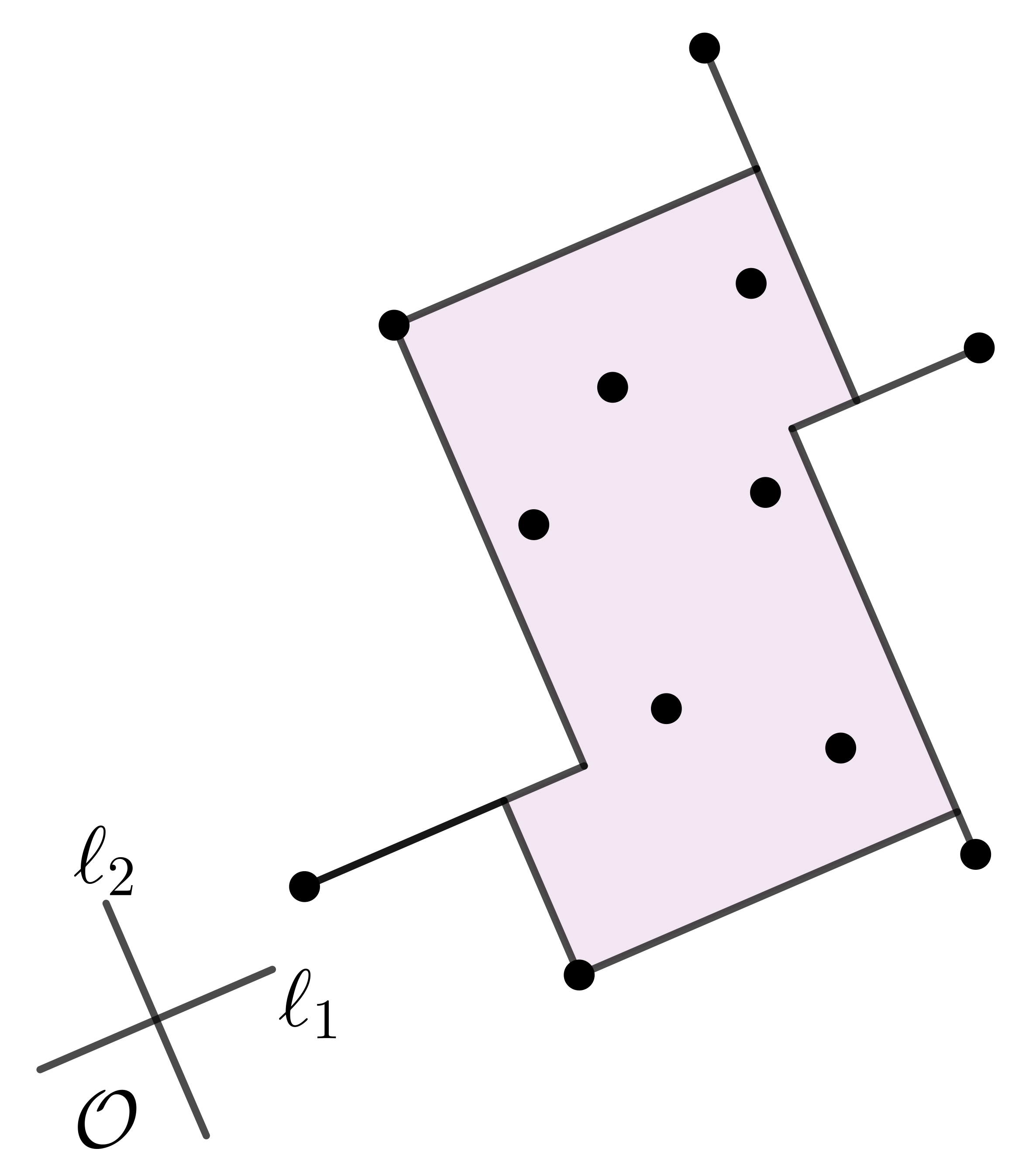}
		\caption{}
		\label{fig_ccc_ohull_2}
	\end{subfigure}
    ~
	\begin{subfigure}[t]{0.3\linewidth}
		\includegraphics[width=\linewidth] {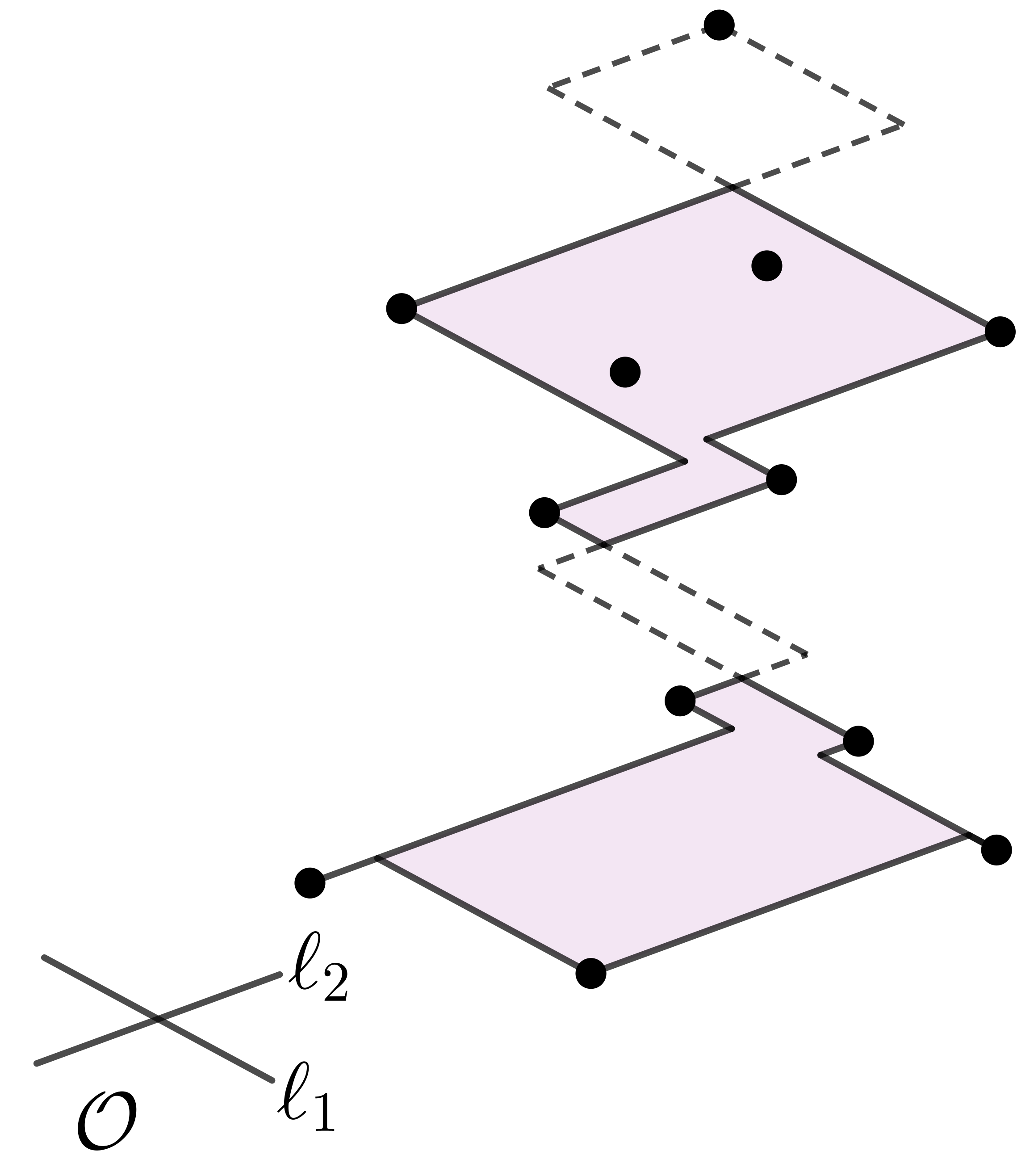}
		\caption{}
		\label{fig_ccc_ohull_3}
	\end{subfigure}
	
	\caption{A point set, its convex hull, and its $\mathcal{O}$-convex hull for two distinct sets of two orientations.}
	\label{fig_ccc_ohull}
\end{figure}

This concept has been generalized in the context of \emph{restricted-orientation geometry}, in which the studied geometric objects comply with properties related to a {f}ixed set of orientations.
Given a set $\mathcal{O}$ of lines through the origin, a set is called $\mathcal{O}$-\emph{convex} if its intersection with any line parallel to a line in $\mathcal{O}$ is either empty or connected\footnote{This generalization of convexity is known as \emph{restricted-orientation convexity}~\cite{fink2004restricted}, \emph{D-convexity}~\cite{matouvsek1998functional}, or \emph{partial convexity}~\cite{naidenko2004partial}.}.
The $\O$-\emph{convex hull} of a geometric set $S$ is de{f}ined as the intersection of all the closed $\O$-convex sets that contain $S$~\cite{fink2004restricted}.
See Figures~\ref{fig_ccc_ohull_2} and \ref{fig_ccc_ohull_3} for illustrations.

An $\O$-convex set is said to be \emph{ortho-convex} in the particular case in which the set $\mathcal{O}$ contains only the coordinate axes; additionally, in this case the $\O$-convex hull of a geometric set $S$ is called its \emph{rectilinear convex hull} (also referred to as orthogonal convex hull) and denoted as $\RH(S)$.

\begin{figure}[ht!]
	\centering
	\begin{subfigure}[t]{0.25\linewidth}
		\includegraphics[width=\linewidth] {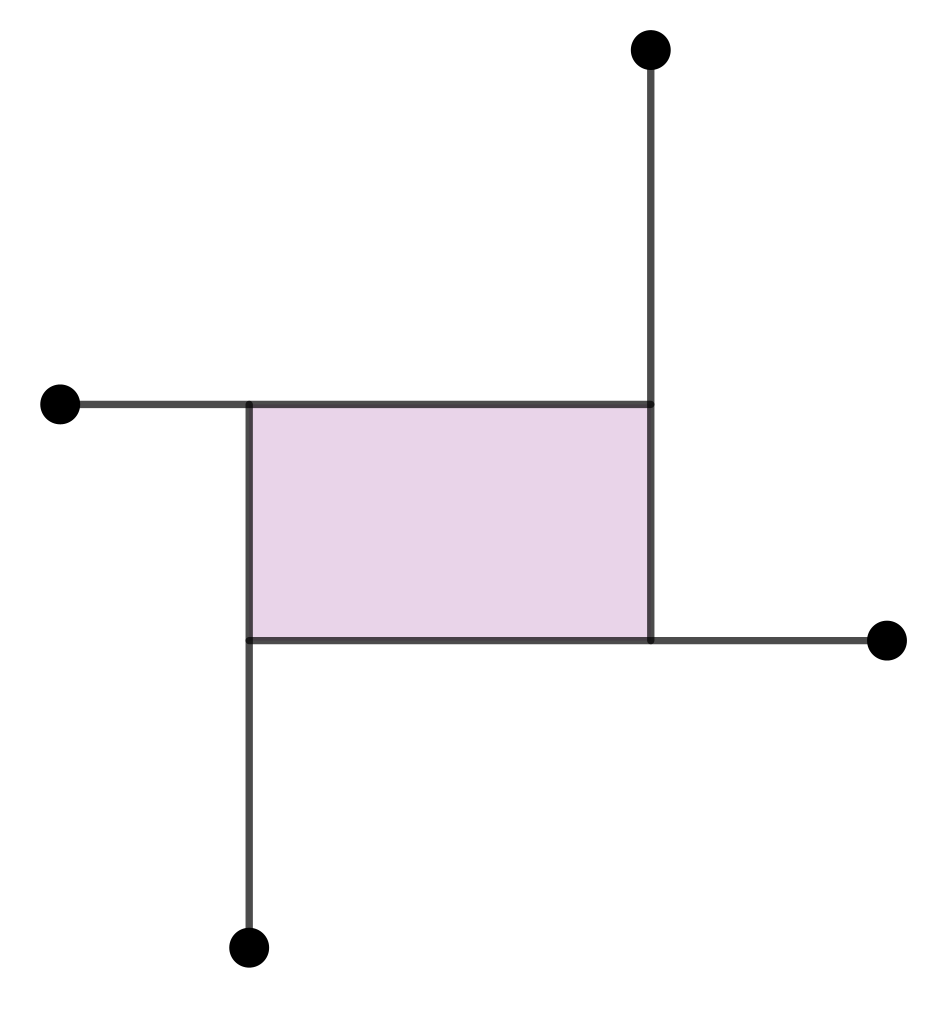}
		\caption{The rectilinear convex hull is connected and has positive area.}
		\label{fig_ccc_1_1}
	\end{subfigure}
	~ 
	\begin{subfigure}[t]{0.25\linewidth}
		\includegraphics[width=\linewidth] {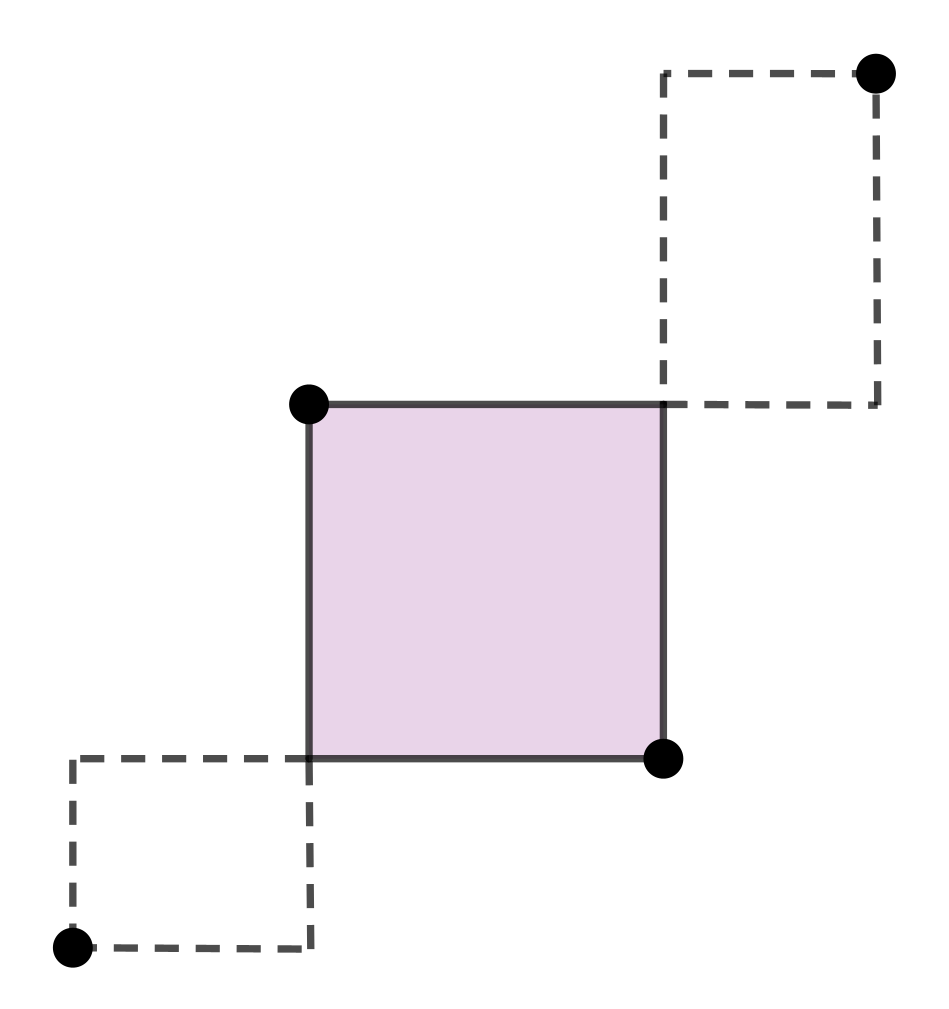}
		\caption{The rectilinear convex hull is not connected and has positive area.}
		\label{fig_ccc_1_2}
	\end{subfigure}
    ~ 
	\begin{subfigure}[t]{0.25\linewidth}
		\includegraphics[width=\linewidth] {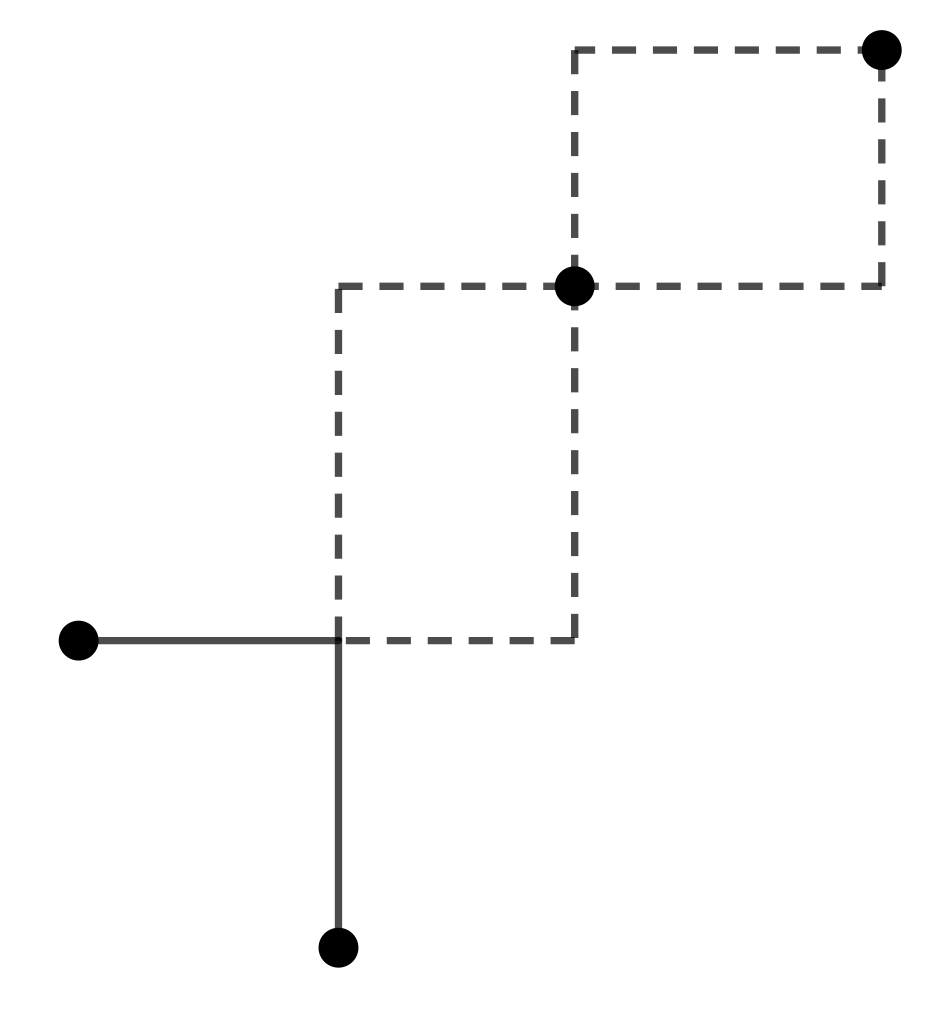}
		\caption{The rectilinear convex hull has zero area.}
		\label{fig_ccc_1_3}
	\end{subfigure}
	
	\caption{The rectilinear convex hull of three sets of four points.}
	\label{fig_ccc_1}
\end{figure}

The rectilinear convex hull $\RH(P)$ of a point set $P$ on the plane can also be de{f}ined in terms of a set of empty open quadrants.
An \emph{open quadrant} is the intersection of two open half-planes whose supporting lines are parallel to the $x$ and $y$ axes, and its \emph{apex} $p$ is the intersection point of these supporting lines.
Given a point set $P$, an open quadrant is said to be \emph{empty} if it contains no elements of $P$.
The rectilinear convex hull $\RH(P)$ of~$P$ is then the region obtained by removing from $\mathbb{R}^2$ all the empty open quadrants; see Figure~\ref{fig_ccc_1} for some examples.
Note that there exist alternative de{f}initions for the $\O$-convex hull of a point set $P$~\cite{ottmann1984definition}.
The de{f}inition of $\RH(P)$ in this paper corresponds to the \emph{maximal $r$-convex hull} of $P$ de{f}ined in~\cite{ottmann1984definition}.
Under this de{f}inition, the rectilinear convex hull of a point set could be disconnected; examples of this are shown in Figures~\ref{fig_ccc_1_2} and~\ref{fig_ccc_1_3}.

\subsection{Related work}

It is known that the rectilinear convex hull of a point set $P$ of size $n$ can be found in $O(n \log n)$ time (see \cite{ottmann1984definition}), and several problems have been studied around this concept.
In \cite{bae2009computing}, Bae \emph{et al.} studied the problem of {f}inding an orientation such that the area of $\RH(P)$ is minimized.
They presented an $O(n^2)$-time algorithm to solve this problem, which can also be used to maintain the rectilinear convex hull of $P$ while rotating the coordinate system.
Later, Diaz-Bañez \emph{et al.} \cite{diaz2011fitting} presented an $O(n \log n)$-time algorithm that can be used to maintain $\RH(P)$ while rotating the coordinate system.
Some of their ideas were subsequently applied by Alegría \emph{et al.}~\cite{alegria2012rectilinear} to compute the orientation minimizing the area of $\RH(P)$ in $O(n \log n)$ time.

The $\O_\beta$-hull of $P$, a generalization of the rectilinear convex hull in which $\O$ consists of two lines forming an angle $\beta$, was studied by Alegría \emph{et al.} in \cite{alegria2018obeta}.
They presented $O(n \log n)$-time algorithms to maintain the $\O_\beta$-hull of $P$, and to {f}ind the angle $\beta$ such that the area or perimeter of the $\O_\beta$-hull of~$P$ is maximized.
Subsequently, Alegría \emph{et al.}~\cite{alegria2021efficient} generalized the previous results to deal with a set of orientations $\O$ consisting of $k \geq 2$ lines; their algorithms are sensitive to two parameters: $k$ and the minimum di{f}{f}erence between $\pi$ and the angle of any of the $2k$ sectors de{f}ined by consecutive lines in $\O$.

Several problems dealing with {f}inding convex subsets of colored point sets have been studied under the classical notion of convexity.
Arévalo \emph{et al.}~\cite{arevalo2022rainbow} studied the problem of deciding if a $4$-colored point set~$P$ of size $n$ in the plane contains a convex \emph{rainbow quadrilateral}: a convex quadrilateral such that its vertices are points in $P$, each of a different color.
They presented an $O(n^2)$-time algorithm to solve this problem, which is equivalent to {f}inding two intersecting segments whose endpoints are points in $P$ having di{f}{f}erent colors.
Bautista \emph{et al.} \cite{bautista2011computing} studied the problem of {f}inding a maximum cardinality \emph{monochromatic island} in a 2-colored point set --which is the intersection of the point set with some convex region, with the property of having points of just one color-- and presented an $O(n^3)$-time algorithm to solve the problem.
Regarding $\O$-convexity, Alegría et al.~\cite{alegria2023separating} studied the problem of computing a set of orientations $\O$ in a 2-colored point set such that the $\O$-convex hull of the points of one color does not contain points of the other color.

Other related problems deal with obtaining partitions of a colored point set into pairwise-disjoint convex subsets complying with some constraints on the colors appearing in each subset.
Holmsen \emph{et al.} \cite{holmsen2017near} proved that for any $d$-colored set $P$ of $n(d + 1)$ points in $\mathbb{R}^d$ in general position with at least $n$ points of each color, there exist $n$ pairwise-disjoint $d$-simplices with vertices in $P$, each of them containing a point of each color.
Problems involving partitions without a {f}ixed number of colors have also been studied.
Van Kreveld \emph{et al.} \cite{van2021diverse} studied the problem of partitioning a colored point set into subsets which are maximally diverse according to two diversity measures, one of them being the sum of the number of di{f}{f}erent colors appearing in each subset.
They proved that the problem of obtaining a diverse partition into Voronoi cells is NP-hard, even when the points lie on the real line, whereas the problem of obtaining a diverse partition into convex subsets is NP-hard in $\mathbb{R}^2$.
Other authors have studied the problem of minimizing the diversity of the subsets to be obtained.
Dumitrescu and Pach~\cite{dumitrescu2002partitioning} proved that any $2$-colored set of $n$ points in general position in~$\mathbb{R}^2$ can be partitioned into at most $\lceil (n+1)/2 \rceil$ monochromatic subsets with pairwise-disjoint convex hulls, and that this bound cannot be improved.
They also proposed an $O(n \log n)$-time algorithm to obtain a partition with minimum size, and provided bounds on the size of such partitions for $k$-colored sets with $k \geq 2$, either in the plane or in higher dimensions.
In an orthogonal setting, Majumder \emph{et al.} \cite{majumder2010separating} proved separating the points of a $k$-colored point set, $k \geq 2$, into monochromatic cells using a minimum number of axis-parallel lines is NP-hard.

\subsection{Problem statement}

Let $P$ be a $k$-colored point set of size $n$ in the plane (without assuming general position).
In this work, we study a variant of the problem of {f}inding a rainbow quadrilateral restricted to the ortho-convex setting: deciding whether $P$ contains a set of four points with di{f}{f}erent colors whose rectilinear convex hull has positive area; see Figures~\ref{fig_ccc_1_1} and \ref{fig_ccc_1_2} for instances of four points whose rectilinear convex hulls have positive area, and Figure~\ref{fig_ccc_1_3} for an example of four points whose rectilinear convex hull has zero area.
Proposition~1 in~\cite{alegria2023separating} and Observation~1 in~\cite{alegria2018obeta} state that a point $x \in \mathbb{R}^2$ is contained in the interior of $\RH(P)$ if and only if every open quadrant with vertex on $x$ contains at least one point of $P$.
We extend this to obtain a region with positive area containing $x$:

\begin{lemma}
\label{lem:open_quadrant}
The rectilinear convex hull of a set $P$ of four points in the plane has positive area if and only if there is a point $p$ such that each of its open quadrants contains a point of $P$.
\end{lemma}
\begin{proof}
\fbox{$\Rightarrow$} Let $p$ be a point in the interior of $\RH(P)$.
Then, each of the four vertices of $\RH(P)$ lies in one of the four quadrants de{f}ined by~$p$ as otherwise $p$ would lie on an empty open quadrant.
This result also follows directly from Proposition~1 in~\cite{alegria2023separating} or Observation~1 in~\cite{alegria2018obeta}.

\fbox{$\Leftarrow$} Let $p$ be a point such that each of its open quadrants contains a point of $P$.
By Proposition~1 in~\cite{alegria2023separating} or Observation~1 in~\cite{alegria2018obeta}, $p$ is in the interior of $\RH(P)$.
Let $B$ be an open ball of radius $\epsilon$ centered at $p$.
Note that every point $b$ of $B$ has the property that each of its open quadrants contains a point of~$P$. This implies that $B$ is completely contained in the rectilinear convex hull of~$P$. The result follows.
\end{proof}

It follows, from the previous observation, that a point set contains a subset of four elements whose rectilinear convex hull has positive area if and only if such four points can be separated by a cross de{f}ined by a vertical and a horizontal line.

\begin{figure*}[!ht]
    \begin{center}
        \includegraphics[width=0.55\linewidth]{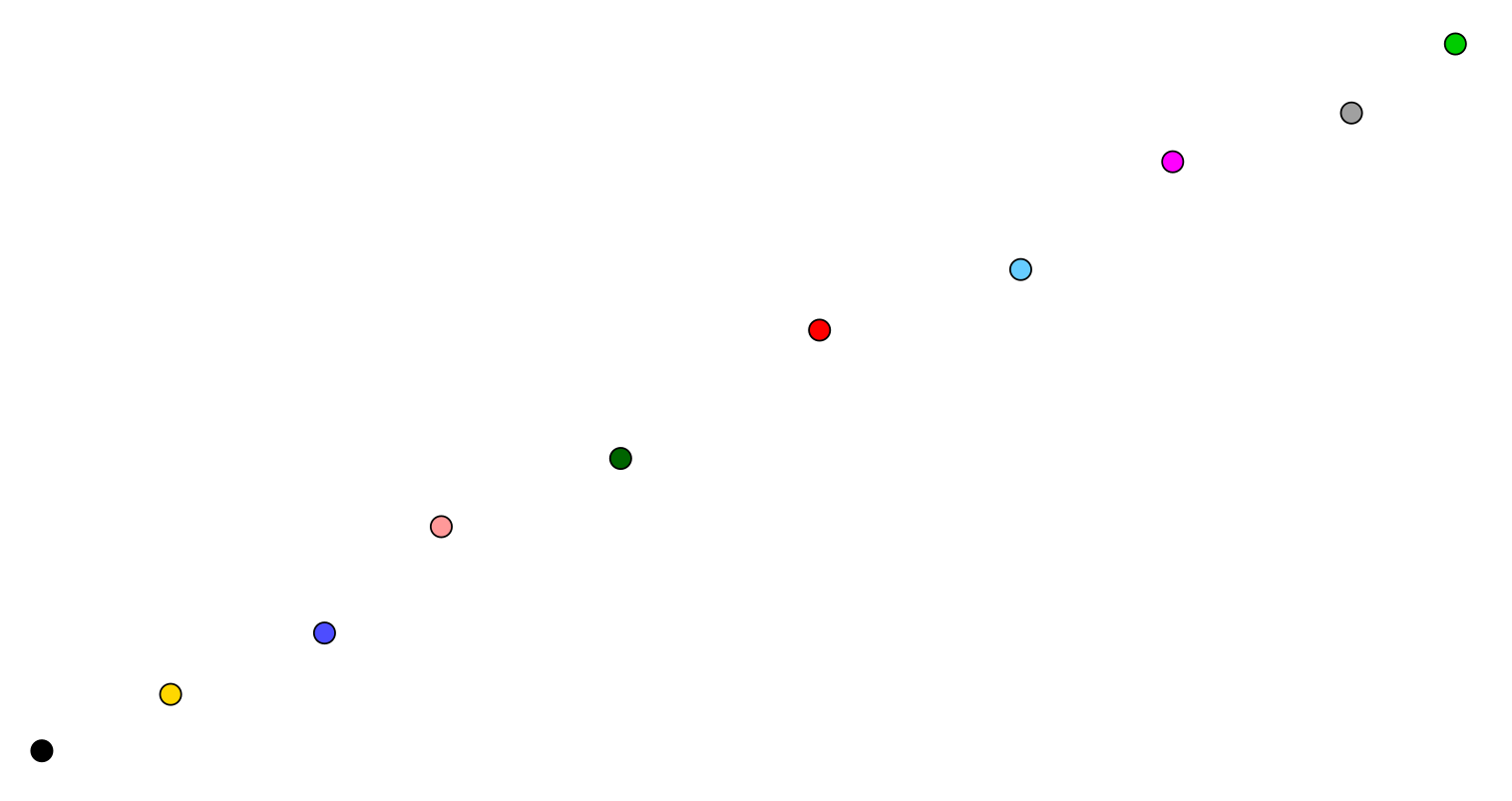}
    \end{center}
    \caption{An $(n-2)$-colored point set of size $n$ containing no subset of four points with di{f}{f}erent colors whose rectilinear convex hull has positive area. Each point is labeled with its color.}
    \label{fig_ccc_2}
\end{figure*}

In contrast with the problem studied in \cite{arevalo2022rainbow}, in which the authors restrict $P$ to be $4$-colored (as any set with $5$ points contains a convex quadrilateral), the ortho-convex setting allows us to de{f}ine our problem for any number $k$ of colors such that $k \leq n$: there are instances of $k$-colored point sets with arbitrary $k \leq n$ for which there exists no subset of four points with di{f}{f}erent colors whose rectilinear convex hull has positive area.
For an example, see Figure~\ref{fig_ccc_2}.

The problem studied in this paper is thus stated as follows:

\begin{prob}[Rainbow Ortho-Convex Positive Area 4-Set Problem]
\label{prob:rocpa4sp}
Let $P$ be a set of size $n$ of $k$-colored points in the plane.
Decide if $P$ contains a subset of four points with di{f}{f}erent colors, whose rectilinear convex hull has positive area.
\end{prob}

Lemma~\ref{lem:open_quadrant} allows us to restate this problem as follows:

\begin{prob}[4-Colored Cross Problem (4CC)]
\label{prob:4ccp}
Let $P$ be a set of size $n$ of $k$-colored points in the plane.
Decide if there exists a point $c$ such that each of the open quadrants of $c$ contains a point of $P$, each of them having a di{f}{f}erent color.
\end{prob}

Our main motivation for the introduction of Problem~\ref{prob:4ccp} is that it will allow us to reason about Problem~\ref{prob:rocpa4sp} in simpler terms, thereby making the proposed solution easier to understand.
On the other hand, Problem~\ref{prob:4ccp} relates Problem~\ref{prob:rocpa4sp} to separability problems on colored point sets using constant-complexity separators~\cite{glazenburg2024robust,houle1993algorithms, hurtado2004separability, hurtado2001separating, hurtado2003red}.

Let $P$ be a set of $k$-colored points in the plane.
If there exists a point~$c$ such that each of its open quadrants contains a point of $P$, each of them having a di{f}{f}erent color, we say that a \emph{4-colored cross} $C$ exists in $P$. In this case, we de{f}ine the point $c$ as the \emph{center} of $C$, and we de{f}ine the four aforementioned points of di{f}{f}erent colors as \emph{witness points} of $C$. Note that a given cross may have more than one set of witness points.

The paper is organized as follows. In section~\ref{sec:upper} we present an $O(n \log n)$-time algorithm for solving the 4CC problem, where the hidden constant is independent of $k$.
In section~\ref{sec:lower} we provide an $\Omega(n \log n)$-time lower bound for the 4CC problem in the algebraic computation tree model. To prove this lower bound we de{f}ine two additional problems: the 2-Colored Open Unitary Gap Problem, and the 2-Colored Negative Slope Problem.
We show that the former one has time complexity $\Omega(n \log n)$ in the algebraic computation tree model using the well-known result by Ben-Or~\cite{ben1983lower}.
The lower bound is then transferred to the 4CC problem by proving a sequence of linear-time reductions, using the 2-Colored Negative Slope Problem as an intermediate problem.

\section{Upper bound}
\label{sec:upper}

We now present an $O(n \log n)$-time algorithm for the 4CC problem, whose running time does not depend on the number of colors $k$.

First, we de{f}ine two useful auxiliary constructs.
\begin{enumerate}
    \item Let $Y=\{y_1,\ldots,y_m\}$ be the set of $y$-coordinates of the points in~$P$, sorted in descending order.
    \item Let $h_i$ be the horizontal line with equation $y=(y_i+y_{i+1})/2$. We de{f}ine the set $H=\{h_1,\ldots, h_{m-1}\}$ as the set of \emph{intermediate} lines.
\end{enumerate}

Our strategy consists of the following three stages:
\begin{enumerate}
    \item Compute the sets $Y$ and $H$.
    \item For each $h_i \in H$, {f}ind four non-empty sets of candidate points, $\NE_i$, $\NW_i$, $\SW_i$, and $\SE_i$, each of them having at most four elements of di{f}{f}erent colors that are extreme in the horizontal direction in each of the four quadrants, respectively.
    For example, $\NE_i$, contains the (at most four) rightmost points of di{f}{f}erent color above $h_i$.
    The {f}irst such point $l_1$ is a point above~$h_i$ with maximal $x$-coordinate; inductively, the $j$-th point~$l_j$ is a point above $h_i$ with maximal $x$-coordinate, after removing from~$P$ the points that share a color with $l_1,\ldots,l_{j-1}$.
    Note that we refer to \emph{a point with the maximal $x$-coordinate} instead of \emph{the rightmost point}, as coordinate values may repeat.
    The other three sets are de{f}ined similarly.
    Notice that $\NE_i$ and $\NW_i$ (resp. $\SW_i$ and $\SE_i$) are not necessarily distinct or disjoint.

	\item For each $h_i \in H$, search for the existence of a 4-colored cross with center on $h_i$, with witness points from $\NE_i, \NW_i, \SW_i$, and $\SE_i$.
\end{enumerate}

The following lemmas imply the correctness of our strategy.

\begin{lemma}
    \label{lem:sepline}
    Let $C$ be a 4-colored cross with witness points $p,q,r$, and $s$ in open quadrants 1 through 4, respectively, of the center $c$ of $C$. Then, there exists an intermediate line $h_i$ which strictly separates $p$ and $q$ from $r$ and $s$.
\end{lemma}
\begin{proof}
    Let $h_c$ be the horizontal line through $c$.
    Since $p$ and $q$ are in the {f}irst two open quadrants of $c$, they lie above $h_c$.
    Similarly, $r$ and $s$ lie below $h_c$.
    Let $y^+$ be the minimum among the $y$-coordinate values of $p$ and $q$ and $y^-$ the maximum among the $y$-coordinate values of $r$ and $s$.
    It follows that $y^+ \neq y^-$, and these two values are then two di{f}{f}erent elements of the set $Y$.
    Therefore, by the de{f}inition of $H$, there exists at least one line in $H$ separating every point with $y$-coordinate at least $y^+$ from every point with $y$-coordinate at most $y^-$.
    The result follows.
\end{proof}

 \begin{lemma}
        \label{lem:sets}
        Let $C$ be a 4-colored cross with witness points $p$, $q$, $r$, and $s$ for the respective four open quadrants of its center $c$. Let $h_i$ be an intermediate line separating $p$ and $q$ from $r$ and $s$, whose existence is guaranteed 
        by Lemma~\ref{lem:sepline}.
        Then, there exists a 4-colored cross with center on $h_i$ each of whose 
        witness points is contained in the corresponding set of candidate points of $h_i$.
\end{lemma}
    \begin{proof}
    Let $d$ be the projection of
     $c$ on $h_i$.
     Since $h_i$ separates $p$ and $q$ from $r$ and $s$, and since the $x$-coordinate of $c$ and $d$ is the same, $p,q,r$ and~$s$ are in the respective {f}irst to fourth open quadrants of $d$. Thus, $d$ is the center of a cross $D$.

     Now, we show how to witness $D$ using only points of the sets of candidate points of $h_i$.

    	Let $p$ be the witness point of $D$ on the {f}irst open quadrant of $d$.
    	If $p \in \NE_i$, leave it as a witness.
    	Otherwise, if there is a point $p'$ of the same color as $p$ in $\NE_i$, use $p'$ as a witness instead of $p$.
    	In the remaining case, there is no point having the same color as $p$ in $\NE_i$, which implies that $\NE_i$ has four elements.
    	Hence, there is at least one point $p' \in \NE_i$ having color di{f}{f}erent from those of the witnesses of $D$ in the other three open quadrants of~$d$, and we can use $p'$ as a witness instead of$p$.
        Notice that the de{f}inition of $\NE_i$ implies that if $p$ is replaced by $p'$ as a witness, then the $x$-coordinate of $p'$ is at least as large as that of $p$. Thus, replaced or not, the witness point is a point in $\NE_i$, it remains in the {f}irst open quadrant of $d$, and its color is di{f}{f}erent from the colors of the other three witnesses.

        Proceeding similarly for the witness points on the remaining open quadrants of $d$, we obtain a set of witness points for $D$, each of its elements contained in the corresponding set of candidate points for $h_i$, as desired.
    \end{proof}

We present the algorithm comprising this strategy as Algorithm~\ref{alg:4cc}.

\begin{algorithm}
    \caption{Four-Colored Cross' algorithm}
    \label{alg:4cc}
    \begin{algorithmic}[1]
    \Require {$P$ is a $k$-colored point set of size $n$ in $\mathbb{R}^2$}
    \Procedure{FourColoredCross}{$P$}
        \State {Sort $P$ by non-decreasing $y$-coordinate of its elements}
        \State {Compute the sets $Y$ and $H$}
        \State {Set $m = \lvert Y \rvert$, $j = 1$, $\NE_1 = \varnothing$, and $\NW_1 = \varnothing$}
        \For{$i = 1, 2, \ldots n$}
            \If{$p_i$ is above $h_j$}
                \State {\Call{TestPoint}{$p_i$, $\NW_j$, $\NE_j$}}
            \Else
                \If{$j < m-1$}
                    \State {Set $j = j + 1$}
                    \State {Set $\NE_j = \NE_{j-1}$ and $\NW_j = \NW_{j-1}$}
                    \State {\Call{TestPoint}{$p_i$, $\NW_j$, $\NE_j$}}
                \EndIf
            \EndIf
        \EndFor
        \State {Set $j = m-1$, $\SE_{m-1} = \varnothing$, and $\SW_{m-1} = \varnothing$}
        \For{$i = n, n-1, \ldots 1$}
            \If{$p_i$ is below $h_j$}
                \State {\Call{TestPoint}{$p_i$, $\SW_j$, $\SE_j$}}
            \Else
                \If{$j > 1$}
                    \State {Set $j = j - 1$}
                    \State {Set $\SE_j = \SE_{j+1}$ and $\SW_j = \SW_{j+1}$}
                    \State {\Call{TestPoint}{$p_i$, $\SW_j$, $\SE_j$}}
                \EndIf
            \EndIf
        \EndFor
        \For{$j = 1, 2, \ldots m-1$}
            \If {\Call{TestCross}{$\NE_j$, $\NW_j$, $\SW_j$, $\SE_j$}}
                \State  \Return {\texttt{true}}
            \EndIf
        \EndFor
        \State \Return {\texttt{false}}
    \EndProcedure
    \end{algorithmic}
\end{algorithm}

\begin{algorithm}
    \caption{Procedure to test a point against candidate sets}
    \label{alg:4cc2}
    \begin{algorithmic}[1]
    \Procedure{TestPoint}{$p$, $W$, $E$}
        \If{$W$ contains a point $q$ of color $c(p)$}
            \If{$q$ is to the right of $p$}
                \State {Replace $q$ with $p$ in $W$}
            \EndIf
        \ElsIf{$W$ has less than $4$ elements}
            \State {Add $p$ to $W$}
        \ElsIf{the rightmost point $q$ of $W$ is to the right of $p$}
            \State {Replace $q$ with $p$ in $W$}
        \EndIf
        \If{$E$ contains a point $q$ of color $c(p)$}
            \If{$q$ is to the left of $p$}
                \State {Replace $q$ with $p$ in $E$}
            \EndIf
        \ElsIf{$E$ has less than $4$ elements}
            \State {Add $p$ to $E$}
        \ElsIf{the leftmost point $q$ of $E$ is to the left of $p$}
            \State {Replace $q$ with $p$ in $E$}
        \EndIf
    \EndProcedure
    \end{algorithmic}
\end{algorithm}

\begin{algorithm}
    \caption{Procedure to test the candidate sets of an intermediate line}
    \label{alg:4cc3}
    \begin{algorithmic}[1]
    \Procedure{TestCross}{$\NE$, $\NW$, $\SW$, $\SE$}
        \ForAll{$p \in \NE$, $q \in \NW$, $r \in \SW$, and $s \in \SE$}
            \If{$p$, $q$, $r$, and $s$ have pairwise di{f}{f}erent colors}
                \If{the open intervals $(q.x, p.x)$ and $(r.x, s.x)$ are not disjoint}
                    \State \Output {$\{p, q, r, s\}$ is a $4$-colored cross}
                    \State \Return {\texttt{true}}
                \EndIf
            \EndIf
        \EndFor
        \State \Return {\texttt{false}}
    \EndProcedure

    \end{algorithmic}
\end{algorithm}


\begin{lemma}
    \label{lem:alg_corr}
    Algorithm~\ref{alg:4cc} returns \texttt{true} if and only if $P$ contains a $4$-colored cross.
\end{lemma}
\begin{proof}
    \fbox{$\Rightarrow$} On the one hand, suppose there exists a 4-colored cross $C$ in~$P$ with center $c$.
    By Lemma~\ref{lem:sepline}, there exists an intermediate line $h_k$ that separates its upper witness points from its lower witness points.
    Moreover, according to Lemma~\ref{lem:sets} there exists a cross with center on $h_k$ which can be witnessed using exactly one point of each set of candidate points of $h_k$.

    Our algorithm sweeps all the points in $P$ in non-decreasing order according to their $y$-coordinate.
    Starting with $h_1$, the algorithm checks if the current point $p_i$ lies above the current intermediate line $h_j$, and if so, it takes the following intermediate line $h_{j+1}$, if such a line exists.
    Thus, the algorithm processes the intermediate lines exhaustively in an ordered way.

    Notice that if a point $p \in P$ is discarded from being in a candidate point set, say $\NE_{i}$, then by an argument similar to the proof of Lemma~\ref{lem:sets}, it can be proven that $p$ does not need to be considered for any candidate point set $\NE_{j}$, with $j > i$.
    That is, either $\NE_{j}$ already contains a point of the same color of $p$ whose $x$-coordinate is not smaller than that of $p$, or $\NE_{j}$ contains four points of di{f}{f}erent colors none of which is to the left of $p$.

    Hence, given the sets of candidate points $\NW_{j-1}$ and $\NE_{j-1}$, we only need to test the points between $h_{j-1}$ and $h_j$ against their elements to obtain $\NW_j$ and $\NE_j$, and each such point is exhaustively tested before $h_{j+1}$ is considered.
    Similarly, each point between $h_{j+1}$ and $h_j$ is tested to obtain $\SW_j$ and $\SE_j$ from $\SW_{j+1}$ and $\SE_{j+1}$.
    Therefore, our algorithm correctly computes the four sets of candidate points associated to each line in $H$.

    Lastly, since for each $h_j \in H$ we test all possible $4$-sets containing one point in each candidate set associated to $h_j$, our algorithm detects a 4-colored cross with center on $h_k$ containing an element in each of the sets $\NW_{k}$, $\NE_{k}$, $\SW_{k}$, and $\SE_{k}$.

    \fbox{$\Leftarrow$} On the other hand, if no 4-colored cross was detected using the sets of candidate points of each and every $h_i \in H$, then, by contraposition of Lemma \ref{lem:sets}, no 4-colored cross exists in $P$.
\end{proof}

We now prove that our strategy can be implemented in $O(n \log n)$ time.

\begin{lemma}
    \label{lem:alg_comp}
    Algorithm~\ref{alg:4cc} runs in $O(n \log n)$ time for an input of $n$ points.
\end{lemma}
\begin{proof}

    The elements of $P$ are {f}irst sorted in $O(n \log n)$ time, after which the sets $Y$ and $H$ can be computed in additional $O(n)$ time.
    Then, the sorted set $P$ is swept twice to obtain the sets of candidate points associated to all the intermediate lines, during which the procedure \textsc{TestPoint} is called at most once per iteration.
    The {f}inal stage calls procedure \textsc{TestCross} once for each of the at most $n-1$ intermediate lines.

    Since the sets of candidate points have size at most four, each call to \textsc{TestPoint} or \textsc{TestCross} takes $O(1)$ time: in the former case, we compare a point against at most four points and update a constant size data structure per candidate set, and in the latter, we test at most $4^4$ subsets of four points.
    Hence, every stage after sorting $P$ takes $O(n)$ time, which implies that Algorithm~\ref{alg:4cc} runs in $O(n \log n)$ time.
\end{proof}

Putting together lemmas~\ref{lem:alg_corr} and~\ref{lem:alg_comp} we obtain the main result of this section:

\begin{theorem} There is an $O(n \log n)$-time algorithm to decide if a $k$-colored point set $P$ of size $n$ contains a 4-colored cross.
\end{theorem}

\section{Lower bound}
\label{sec:lower}

We {f}irst de{f}ine the following auxiliary problems:

\begin{prob}[2-Colored Open Unitary Gap Problem (2COUG)]
Given two sets of real numbers $\{x_1, \ldots, x_n\}$ and $\{y_1, \ldots, y_n\}$, decide if there exists a pair $x_i, y_j$ such that $0 < |x_i-y_j| < 1$.
\end{prob}

\begin{prob}[2-Colored Negative Slope Problem (2CNS)]
Given a 2-colored set $P$ of $n$ points in the plane, decide if there is a pair of points of di{f}{f}erent colors in $P$ such that the line passing through them has negative slope.
\end{prob}

We {f}irst prove the lower bound on the time complexity of the 2COUG problem in the algebraic computation tree model.
Our proof follows the same ideas as the original proof by Ben-Or~\cite{ben1983lower} of the lower bound for the Element Distinctness Problem, which consists in deciding if a sequence with $n$ reals contains two elements with the same value.
Given a set $W \subseteq \mathbb{R}^n$, the Membership Problem for $W$ is that of determining if a given $x = (x_1, \ldots, x_n) \in \mathbb{R}^n$ is contained in $W$.
By setting $W = \{(x_1, \ldots , x_n) \mid \prod_{i \neq j}(x_i - x_j) \neq 0 \}$, the problem of deciding if a given $x = (x_1, \ldots, x_n)$ is contained in $W$ is equivalent to the element distinctness problem on the sequence $[x_1, \ldots, x_n]$.

\begin{theorem}[Ben-Or~\cite{ben1983lower}]
    \label{thm:ben-or}
    Let $W \subseteq \mathbb{R}^n$ be any set. The complexity of the membership problem on $W$ in the algebraic computation tree model is $\Omega(\log N - n)$, where $N$ is the maximum between the number of connected components in $W$ and the number of connected components in $\mathbb{R}^n - W$.
\end{theorem}

By noting that each region $\{(x_1, \ldots , x_n) \mid x_{\sigma(1)} < x_{\sigma(2)} < \ldots < x_{\sigma(n)} \}$ is a maximal connected component of $W$ for each permutation $\sigma$, Ben-Or obtained the following result:

\begin{theorem}[Ben-Or~\cite{ben1983lower}]
    \label{thm:ben-or2}
    Any algebraic computation tree that solves the $n$-element distinctness problem must have complexity of at least $\Omega({n \log n})$.
\end{theorem}

Now we demonstrate that the 2-Colored Open Unitary Gap Problem is equivalent to solving the membership problem for a set $W$ with $(n!)^2$ connected components.

\begin{lemma}
    The 2-Colored Open Unitary Gap Problem has time complexity $\Omega(n \log n)$ in the algebraic computation tree model.
\end{lemma}
\begin{proof}
 Consider the set of points in $\mathbb{R}^{2n}$ that correspond to NOT-instances of 2COUG, that is, the set of points $p=(x_1,\ldots,x_n,y_1,\ldots,y_n) = (z_1,\ldots,z_{2n})$ such that for every pair $x_i,y_j$, $|x_i-y_j|$ is not in the open interval $(0,1)$.
 We {f}irst show that the solution space for this problem has at least $(n!)^2$ connected components.

    Let $W$ be the following set of $(n!)^2$ points in $\mathbb{R}^{2n}$:
    \begin{multline*}
        W :=\{\,p_{i,j}= \big(\sigma_i(1),\sigma_i(3),\ldots,\sigma_i(2n-1),\\
        \pi_j(2),\pi_j(4),\ldots,\pi_j(2n)\big) \, | \, i,j \in [n!] \,\}
    \end{multline*}

    In other words, the {f}irst $n$ coordinates of the point $p_{i,j}$ are the permutation $\sigma_i$ of the sequence $[1,3,\ldots,2n-1]$, and its last $n$ coordinates are the permutation $\pi_j$ of the sequence $[2,4,\ldots,2n]$.

    Clearly, every point of $W$ corresponds to a NOT-instance of 2COUG.
    We claim that any two distinct points $s=p_{i,j} \neq p_{i',j'}=t$ in $W$ are contained in separate connected components.
    Suppose w.l.o.g. that $i \neq i'$.
    Let $z_l(p)$ denote the $l$-th coordinate of point $p$.
    Then, there exist two distinct indices $a,b \in [1,n]$, such that
    $z_a(s) < z_b(s)$ and $z_a(t) > z_b(t)$.
    Note that there also exist an even integer $e$, and an index $c \in [n+1,2n]$,  such that $z_a(s) < z_c(s)=e < z_b(s)$.

    Let $\alpha$ be a continuous path from $s$ to $t$ in $\mathbb{R}^{2n}$. Since  $z_b(t) < z_a(t)$, then either $z_c(t) < z_a(t)$, or $z_b(t) < z_c(t)$. In any case, when moving continuously along $\alpha$ from $s$ to $t$, the value of the $c$-th coordinate changes its relative ordering with the  value of the $a$-th coordinate, or it changes its relative ordering  with the value of the $b$-th coordinate. Thus, the distance between the two involved values takes every value in the open interval $(0,1)$, which corresponds to passing through points corresponding to YES-instances of 2COUG. This proves our claim that $s$ and $t$ are in separate connected components.

By applying Theorem~\ref{thm:ben-or}, it follows that any algorithm solving this problem in the algebraic computation tree model has complexity $\Omega(\log(n!^2)) \allowbreak  =\Omega(n \log n)$.
\end{proof}

Now we transfer this lower bound to the 4CC problem by means of two linear time reductions: one from 2COUG to 2CNS, the other from 2CNS to~4CC.

\begin{figure*}[!ht]
    \begin{center}
    \includegraphics[width=0.65\linewidth]{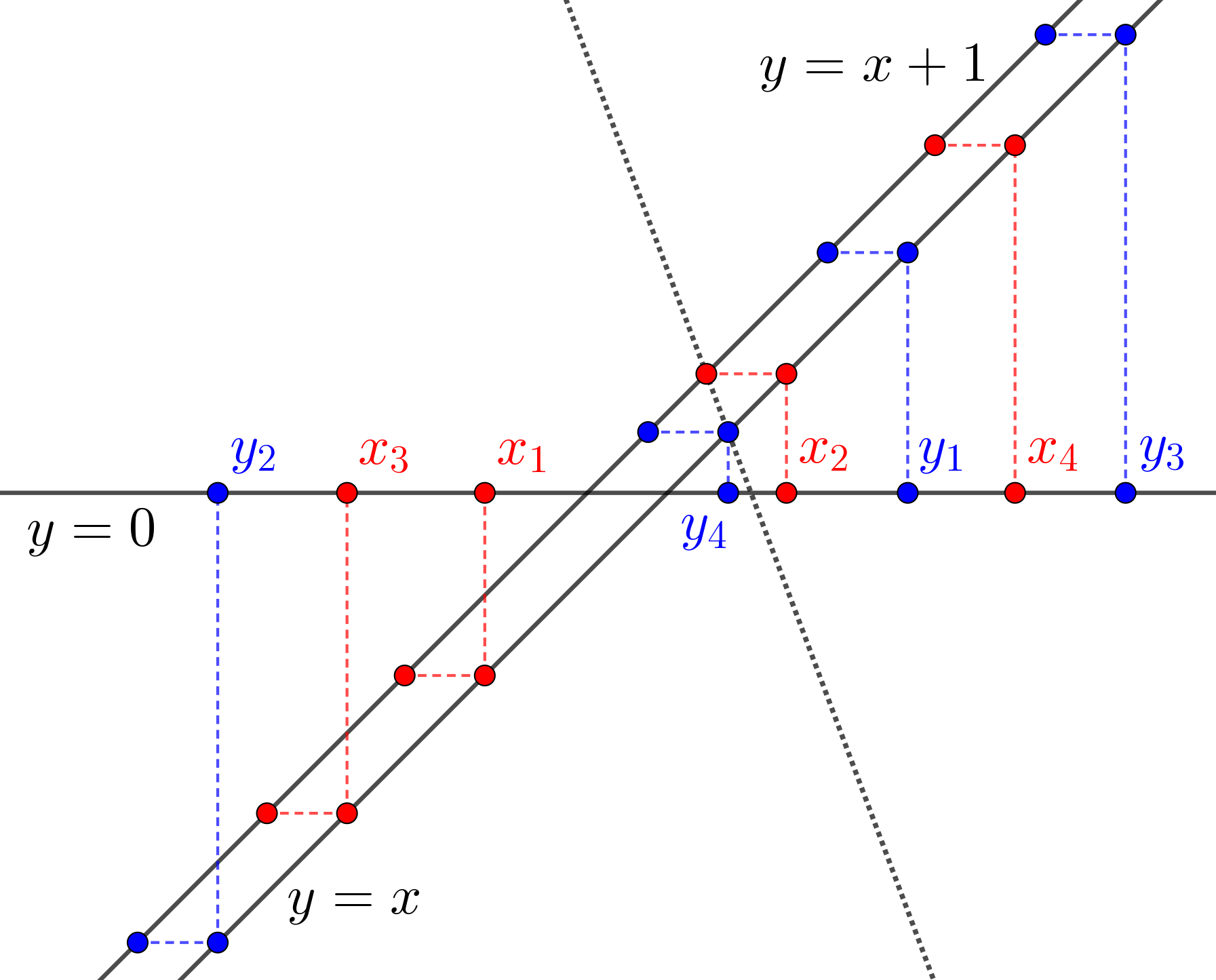}
    \end{center}
    \caption{Illustration of the proof of Lemma~\ref{lem:2couip_2cns}.}
    \label{fig_ccc_4}
\end{figure*}

\begin{lemma}
    \label{lem:2couip_2cns}
    The 2-Colored Open Unitary Gap Problem can be reduced to the 2-Colored Negative Slope Problem in linear time.
\end{lemma}
\begin{proof}
    Given an instance $I=\{x_1,\ldots,x_n,y_1,\ldots,y_n\}$ of the 2COUG problem, construct the following instance~$O$ of the 2CNS problem:

    Let $L$ be a set of $2n$ points on the line $y = x$ containing a red point $(x_i, x_i)$ for each $x_i \in \{x_1,\ldots,x_n\}$, and a blue point $(y_j, y_j)$ for each $y_j \in \{y_1,\ldots,y_n\}$.
    Let $U$ be a set of $2n$ points on the line $y = x+1$ containing, for each point $(a,b) \in L$, a point $(a-1,b)$ having the same color.
    We de{f}ine $O = L \cup U$ as an instance of size $4n$ of the 2CNS problem.
    An example of this construction is shown in Figure~\ref{fig_ccc_4}.

    It is clear that the construction of $O$ takes linear time.
    Now, we prove that $I$ is a YES-instance of 2COUG if and only if $O$ is a YES-instance of 2CNS.

    Assume that $I$ is a YES-instance of 2COUG.
    Let $x_i$, $y_j$ be a pair of entries such that $0<|x_i-y_j|<1$.
    Suppose w.l.o.g. that $y_j < x_i$.
    Then, $O$ contains a red point $(x_i -1, x_i)$ and a blue point $(y_j, y_j)$.
    Note that $(y_j-x_i) < 0$, and that $(y_j-x_i+1)>0$. Thus, the slope of the line through the aforementioned points is negative.
    Therefore, $O$ is a YES-instance of 2CNS.

    Now assume that $O$ is a YES-instance of the 2CNS problem. Let $p, q$ be a pair of points of di{f}{f}erent colors yielding a line with negative slope.
    Observe that exactly one of these points lies on the line $y=x$ and the other one lies on the line $y=x+1$.
    Suppose w.l.o.g. that $p = (z, z)$ is on the line $y = x$.
    Then, $q$ is contained in the open intersection of the second open quadrant of $p$ and the line $y=x+1$. It follows that the coordinates of $q$  are $(z-1+\epsilon,z+\epsilon)$, for some $0 < \epsilon < 1$.

    By construction, $p = (z, z)$ is an element of $L$ corresponding to a value $z \in I$ whereas $q = (z-1+\epsilon,z+\epsilon)$ is an element of $U$ created from a point $(z+\epsilon,z+\epsilon) \in L$ corresponding to a value $z+\epsilon \in I$.
    Furthermore, since $p$ and $q$ have di{f}{f}erent colors, it follows that one was constructed from one of the {f}irst $n$ entries of $I$ while the other was constructed from one of the last $n$ entries of $I$. Thus, $I$ is a YES-instance of 2COUG.
\end{proof}

\begin{lemma}
    The 2-Colored Negative Slope Problem can be reduced to the 4-Colored Cross problem in linear time.
\end{lemma}
\begin{proof}
Let $I$ be an instance of the 2CNS problem consisting of $n$ red/blue points.
    We construct an instance $O$ of 4CC in linear time.

    Let $\min_x$ and $\max_x$ be the minimum and maximum $x$-coordinates, respectively, among the elements in $I$; $\min_y$ and $\max_y$ are de{f}ined similarly for $y$-coordinates.
    Let $r = (\max_x + 1, \max_y + 1)$ be a green point and $s = (\min_x - 1, \min_y - 1)$ be a black point.
    De{f}ine $O = I \cup \{r, s\}$.
    Clearly, $O$ can be obtained in $O(n)$ time.

    Now we show that $I$ is a YES-instance of 2CNS if and only if $O$ is a YES-instance of 4CC.

    Assume that $I$ is a YES-instance of 2CNS.
    Let $p,q$ be a pair of points of di{f}{f}erent colors in  $I$ yielding a line with negative slope.
    Let $c$ be the midpoint of the segment joining $p$ and $q$.
    Note that each of the four open quadrants de{f}ined by $c$ contains a point in $\{p,q,r,s\}$, each of which has a di{f}{f}erent color.
    Thus, $O$ is a YES-instance of 4CC.

    Assume now that $O$ is a YES-instance of 4CC.
    Let $c$ be the center of some 4-colored cross in $O$.
    This cross must include the extreme points $r$ and~$s$, since they are the only green and black points of~$O$; and they must be located in the {f}irst and third open quadrants of $c$, respectively, since any other possibility yields at least two open quadrants of $c$ empty of points of~$O$. Let $p,q$ be the points of $O$ in the second and fourth open quadrants of $c$. Note that $p,q$ are points in $I$, of di{f}{f}erent colors, and they yield a line with negative slope.
    It follows that $I$ is a YES-instance of 2CNS.
\end{proof}

The following result is a consequence of the preceding lemmas:

\begin{theorem}
The 4-Colored Cross Problem has time complexity $\Omega(n \log n)$ in the algebraic computation tree model.
\end{theorem}

The upper and lower bounds for the 4-Colored Cross Problem translate directly into upper and lower bounds for the Rainbow Ortho-Convex Positive Area 4-Set Problem. We thus conclude with the following result:

\begin{theorem}
    The Rainbow Ortho-Convex Positive Area 4-Set Problem can be solved in $O(n\log n)$ time,
    and this time complexity is a lower bound for this problem in the algebraic computation tree model.
\end{theorem}

\section{Conclusion and future research}

In this paper we provided an $O(n \log n)$-time algorithm for the problem of deciding if a $k$-colored point set $P$ of size $n$ in the plane contains a subset of four elements, all of them of di{f}{f}erent colors, whose rectilinear convex hull has positive area.
We also proved an $\Omega(n \log n)$ lower bound for this problem in the algebraic computational tree model.
We achieved this by restating the problem as the equivalent one of deciding if there exists a vertical and a horizontal line such that each of the four open regions generated by its intersection contains a point of $P$ whose color is di{f}{f}erent from those of the other three points.

As per future research directions, a natural extension for the $4$-colored cross problem is the following problem: given a $k$-colored point set $P$, {f}ind a line~$\ell_1$ with {f}ixed orientation (e.g., vertical) and another line $\ell_2$ such that each of the four regions in which these lines divide the plane contains a point of~$P$ with di{f}{f}erent color from those in the other regions and the angle between $\ell_1$ and $\ell_2$ is minimized/maximized.

\section*{Acknowledgments}
We thank Alma Arévalo for her participation in the initial discussions of this problem.

The authors wish to acknowledge and thank various funding sources.
David Flores-Peñaloza was partially supported by grant PAPIIT-IN115923, UNAM, México; Mario A.\ Lopez was partially supported by an Evans Research Fund of the University of Denver; Nestaly~Mar\'in was supported by UNAM Posdoctoral Program (POSDOC) and partially supported by grant PAPIIT-IN115923, UNAM, México; David Orden was supported by Grant PID2019-104129GB-I00 funded by MICIU/AEI/10.13039/501100011033.


\begin{thebibliography}{22}

\bibitem{alegria2021efficient}
C.~Alegr{\'{\i}}a, D.~Orden, C.~Seara, J.~Urrutia,
Efficient computation of minimum-area rectilinear convex hull under rotation and generalizations,
Journal of Global Optimization 79 (2021) 687--714.
\bibitem{alegria2023separating}
C.~Alegr{\'{\i}}a, D.~Orden, C.~Seara, J.~Urrutia,
Separating bichromatic point sets in the plane by restricted orientation convex hulls,
Journal of Global Optimization 85(4) (2023) 1003--1036.
\bibitem{alegria2012rectilinear}
C.~Alegr{\'{\i}}a-Galicia, T.~Garduno, A.~Rosas-Navarrete, C.~Seara, J.~Urrutia,
Rectilinear convex hull with minimum area,
in: Computational Geometry: XIV Spanish Meeting on Computational Geometry, EGC 2011, Dedicated to Ferran Hurtado on the Occasion of His 60th Birthday, Alcal{\'a} de Henares, Spain, June 27-30, 2011, Revised Selected Papers,
Springer, 2012, pp. 226--235.
\bibitem{alegria2018obeta}
C.~Alegr{\'{\i}}a-Galicia, D.~Orden, C.~Seara, J.~Urrutia,
On the {$O_\beta$}-hull of a planar point set,
Computational Geometry 68 (2018) 277--291.
\bibitem{arevalo2022rainbow}
A.~Ar{\'e}valo, R.~Ch{\'a}vez-Jim{\'e}nez, A.~Hern{\'a}ndez-Mora, R.~L{\'o}pez-L{\'o}pez, N.~Mar{\'\i}n, A.~Ram{\'\i}rez-Vigueras, O.~Sol{\'e}-Pi, J.~Urrutia,
On Rainbow Quadrilaterals in Colored Point Sets,
Graphs and Combinatorics 38(5) (2022) 152.
\bibitem{bae2009computing}
S.~W. Bae, C.~Lee, H.-K. Ahn, S.~Choi, K.-Y. Chwa,
Computing minimum-area rectilinear convex hull and {L}-shape,
Computational Geometry 42(9) (2009) 903--912.
\bibitem{bautista2011computing}
C.~Bautista-Santiago, J.~M. D{\'\i}az-B{\'a}{\~n}ez, D.~Lara, P.~P{\'e}rez-Lantero, J.~Urrutia, I.~Ventura,
Computing optimal islands,
Operations Research Letters 39(4) (2011) 246--251.
\bibitem{ben1983lower}
M.~Ben-Or,
Lower bounds for algebraic computation trees,
in: Proceedings of the fifteenth Annual ACM Symposium on Theory of Computing, 1983, pp. 80--86.
\bibitem{diaz2011fitting}
J.~M. D{\'\i}az-B{\'a}{\~n}ez, M.~A. Lopez, M.~Mora, C.~Seara, I.~Ventura,
Fitting a two-joint orthogonal chain to a point set,
Computational Geometry 44(3) (2011) 135--147.
\bibitem{dumitrescu2002partitioning}
A.~Dumitrescu, J.~Pach,
Partitioning colored point sets into monochromatic parts,
International Journal of Computational Geometry \& Applications 12(05) (2002) 401--412.
\bibitem{fink2004restricted}
E.~Fink, D.~Wood,
Restricted-orientation convexity,
Monographs in Theoretical Computer Science (An EATCS Series), Springer, 2004.
\bibitem{glazenburg2024robust}
E.~Glazenburg, T.~van~der Horst, T.~Peters, B.~Speckmann, F.~Staals,
Robust Bichromatic Classification using Two Lines,
arXiv preprint arXiv:2401.02897.
\bibitem{holmsen2017near}
A.~F. Holmsen, J.~Kyn{\v{c}}l, C.~Valculescu,
Near equipartitions of colored point sets,
Computational Geometry 65 (2017) 35--42.
\bibitem{houle1993algorithms}
M.~F. Houle,
Algorithms for weak and wide separation of sets,
Discrete Applied Mathematics 45(2) (1993) 139--159.
\bibitem{hurtado2004separability}
F.~Hurtado, M.~Mora, P.~A. Ramos, C.~Seara,
Separability by two lines and by nearly straight polygonal chains,
Discrete Applied Mathematics 144(1-2) (2004) 110--122.
\bibitem{hurtado2001separating}
F.~Hurtado, M.~Noy, P.~A. Ramos, C.~Seara,
Separating objects in the plane by wedges and strips,
Discrete Applied Mathematics 109(1-2) (2001) 109--138.
\bibitem{hurtado2003red}
F.~Hurtado, C.~Seara, S.~Sethia,
Red-blue separability problems in 3D,
in: International Conference on Computational Science and Its Applications, Springer, 2003, pp. 766--775.
\bibitem{van2021diverse}
M.~van Kreveld, B.~Speckmann, J.~Urhausen,
Diverse partitions of colored points,
in: Algorithms and Data Structures: 17th International Symposium, WADS 2021, Virtual Event, August 9--11, 2021, Proceedings 17, Springer, 2021, pp. 641--654.
\bibitem{majumder2010separating}
S.~Majumder, S.~C. Nandy, B.~B. Bhattacharya,
Separating multi-color points on a plane with fewest axis-parallel lines,
Fundamenta Informaticae 99(3) (2010) 315--324.
\bibitem{matouvsek1998functional}
J.~Matou{\v{s}}ek, P.~Plech{\'a}{\v{c}},
On functional separately convex hulls,
Discrete \& Computational Geometry 19 (1998) 105--130.
\bibitem{naidenko2004partial}
V.~G. Naidenko,
Partial convexity,
Mathematical Notes 75 (2004) 202--212.
\bibitem{ottmann1984definition}
T.~Ottmann, E.~Soisalon-Soininen, D.~Wood,
On the definition and computation of rectilinear convex hulls,
Information Sciences 33(3) (1984) 157--171.


\end{thebibliography}

\end{document}